\documentclass{aa}  
\usepackage{graphicx}
\usepackage{tabularx}
\usepackage{txfonts}
\usepackage{footnote}
\usepackage{soul}

\newcommand{\teff}{$T_{\rm eff}$ }
\newcommand{\tsin}{$T_{\rm eff}$}

\begin{document}

   \title{The Solar Twin Planet Search}

   \subtitle{V. Close-in, low-mass planet candidates and evidence of planet accretion in the solar twin HIP 68468}

      \author{Jorge Mel\'{e}ndez\inst{1}
          \and 
          Megan Bedell\inst{2}
          \and
          Jacob L. Bean\inst{2}
          \and
          Iv\'an Ram{\'{\i}}rez\inst{3}
          \and
          Martin Asplund\inst{4}
          \and
          Stefan Dreizler\inst{5}
          \and
          Hong-Liang Yan\inst{6}
         \and
         Jian-Rong Shi\inst{6}
         \and
          Karin Lind\inst{7}
          \and
          Sylvio Ferraz-Mello\inst{1}
          \and
          Jhon Yana Galarza\inst{1}
          \and
          Leonardo dos Santos\inst{1}
          \and
          Lorenzo Spina\inst{1}
          \and
          Marcelo Tucci Maia\inst{1}
          \and
          Alan Alves-Brito\inst{8}
          \and
          TalaWanda Monroe\inst{9}
          \and
          Luca Casagrande\inst{4}
          }

   \institute{Universidade de S\~ao Paulo, IAG, Departamento de Astronomia,  Rua do Mat\~ao 1226, 
              Cidade Universit\'aria, 05508-900 S\~ao Paulo, SP, Brazil. \email{jorge.melendez@iag.usp.br}
         \and
             University of Chicago, Department of Astronomy and Astrophysics, 5640 S. Ellis Ave, Chicago, IL 60637, USA
         \and
         University of Texas at Austin, McDonald Observatory and Department of Astronomy, USA
         \and
             The Australian National University, Research School of Astronomy and Astrophysics, Cotter Road, Weston, ACT 2611, Australia    
         \and
         University of G\"ottingen, Institut f\"ur Astrophysik, Germany
         \and
         Key Laboratory of Optical Astronomy, National Astronomical Observatories, Chinese Academy of Sciences, Beijing 100012, China
         \and
         Max Planck Institute for Astronomy, K\"onigstuhl 17, 69117 Heidelberg, Germany
         \and
         Universidade Federal do Rio Grande do Sul, Instituto de Fisica, Av. Bento Goncalves 9500, Porto Alegre, RS, Brazil
         \and
         Space Telescope Science Institute, Baltimore, MD, USA}

   \date{Received ... 2015; accepted ... 2016}

 
  \abstract
   {More than two thousand exoplanets have been discovered to date. Of these, only a small fraction 
   have been detected around solar twins, which are key stars because we can 
   obtain accurate elemental abundances especially for them, which is crucial for studying the planet-star chemical connection 
   with the highest precision.}
   {We aim to use solar twins to characterise the relationship between planet architecture and stellar chemical composition.}
   {We obtained high-precision (1 m s$^{-1}$) radial velocities with the HARPS spectrograph on the ESO 3.6 m telescope at La Silla Observatory 
   and determined precise stellar elemental abundances ($\sim$0.01 dex) using spectra obtained with the MIKE spectrograph on the Magellan 6.5m telescope.}
   {Our data indicate the presence of a planet with a minimum mass of 26 $\pm$ 4 Earth masses around the solar twin HIP 68468. The planet is more massive
   than Neptune (17 Earth masses), but unlike the distant Neptune in our solar system
   (30 AU),  HIP 68468c is close-in, with a semi-major axis of 0.66 AU, similar to that of Venus. 
   The data also suggest the presence of a super-Earth with a minimum mass of 2.9 $\pm$ 0.8 Earth masses at 0.03 AU;
   if the planet is confirmed, it will be the fifth least massive radial velocity planet candidate discovery to date and the first super-Earth around a solar twin.
   Both isochrones (5.9$\pm$0.4 Gyr) and the abundance ratio [Y/Mg] (6.4$\pm$0.8 Gyr) indicate an age of about 6 billion years. 
   The star is enhanced in refractory elements when compared to the Sun, and the refractory enrichment is
   even stronger after corrections for Galactic chemical evolution.
   We determined a nonlocal thermodynamic equilibrium Li abundance of 1.52$\pm$0.03 dex, which is four times higher than what would be expected for the
   age of HIP 68468. The older age is also supported by the low log(R'$_{HK}$) (-5.05) and low jitter ($<$1 m s$^{-1}$). 
    Engulfment of a rocky planet of 6 Earth masses can explain the enhancement in both lithium and the refractory elements.   
}
   {The super-Neptune planet candidate is too massive for in
situ formation,
and therefore its current location is most likely the result of planet
migration that could also have driven other planets towards its host star,
enhancing thus the abundance of lithium and refractory elements in HIP
68468. The intriguing evidence of planet accretion warrants further
observations to verify the existence of the planets that are
indicated by our data
and to better constrain the nature of the planetary system around this
unique star. }


   \maketitle
%

\section{Introduction}

Soon after the discovery of the first exoplanets around solar-type 
stars \citep{mq95,bm96,mb96} a connection was noted between 
the metallicity of planet-host stars and the presence of close-in giant planets \citep{gon97}. 
Since then, our techniques to detect planets and to characterise stars 
have been substantially improved. It is now possible to achieve a precision of 1 m s$^{-1}$
in radial velocity (RV), increasing our abilities to detect nearby small planets and distant 
large planets \citep[e.g.][]{san04,how11,bed15}.
Additionally, a precision of 0.01 dex in 
chemical abundances has been achieved in solar twins \citep[e.g.][]{bed14,mel14a,ram14,tuc14,nis15}. 
This is to be compared with a precision of $\sim$10 m s$^{-1}$ for the first planet detections \citep{mq95,bm96,mb96} 
and errors of 0.08-0.09 dex in iron abundances of the first planet hosts \citep{gon97}.

Aiming to explore the planet-star connection more thoroughly, we are exploiting the synergy 
between precise chemical abundances that can be achieved in solar twins (0.01 dex) and precise 
planet characterisation that can be achieved with HARPS (1 m s$^{-1}$) through an on-going 
survey of planets around solar twins.
Our sample was presented in \cite{ram14}, where we determined stellar parameters 
(\tsin, log $g$, [Fe/H]), ages, masses, and stellar activity. 
Slightly improved ages and the [Y/Mg]-age relation were published in \cite{tuc16}.
Our HARPS data have previously been
used to obtain stringent constraints on planets around two important solar twins, 18 Sco, which is the
brightest solar twin and is younger than the Sun \citep{mel14a},
and HIP 102152, an old solar twin  with a low lithium content \citep{mon13}.
Our HARPS data were also used to study the rotation-age relation,
showing that our Sun is a regular rotator \citep{leo16}.

The first planet detected in our HARPS survey is a Jupiter twin around the solar twin HIP 11915, a planet with about the mass of Jupiter 
and at about the same distance as the distance of Jupiter from the Sun \citep{bed15}. Interestingly, the planet host star 
has the same volatile-to-refractory ratio as in the Sun, suggesting that in addition to being a 
Jupiter twin, it may host rocky planets, according to the hypothesis put forward by \cite{mel09}.

In this paper we present the second and third planets discovered from our dedicated solar twin planet search.
The evidence for the outer planet seems solid, but the inner planet is less well constrained.
We discuss how the planet architecture may be related to the chemical abundance pattern.


\section{Planet detection}

\subsection{Data}

\label{data}

HIP 68468 was observed with the HARPS spectrograph \citep{may03} on the ESO 3.6 m telescope at La Silla Observatory during 43 nights between 2012-2016 (the star was observed twice in two nights, resulting in 45 data points). These observations were carried out under our large program targeting 63 solar twin stars (program ID 188.C-0265). Each night's observation consisted of one 1300-second exposure for this $V=9.4$ star. This exposure time is long enough to average out $p$-mode stellar oscillations typical of solar-like stars, and to achieve
a photon-limited precision of 1 m s$^{-1}$ in typical observing conditions. 
On two of the observing nights, two 1300-second exposures were taken with a two-hour time gap between them.  In total, we obtained 45 RV measurements for the star. 

All HARPS data were processed with version 3.8 of the dedicated HARPS pipeline, which determines radial velocities through a cross-correlation technique using a G2 binary mask. Uncertainties on the RVs came from the pipeline estimation, which accounts for photon noise in the cross-correlation. 
An additional 1 m s$^{-1}$ was added in quadrature to account for the instrumental noise floor \citep{Dumusque2011}. The Ca II H\&K spectral lines were measured by us, and the resulting $S_{HK}$ activity indices were converted into the standard Mount Wilson log($R'_{HK}$) index using the relation derived by \citet{Lovis2011}. 

The HARPS instrument underwent an upgrade in June 2015 that included the installation of new fibers and instrumental re-focusing. This upgrade altered the instrumental profile and introduced offsets in the measured RVs relative to the pre-upgrade RV zero point. 
We used the data for a set of ten stars in our program with relatively constant pre-upgrade RV time series (RMS < 2 m s$^{-1}$) to characterise the offset. Although this offset is expected to have a dependence on stellar type, all stars in our program are solar twins and should therefore share the same offset. We find an offset of 15.4 $\pm$ 0.2 m s$^{-1}$ among the constant solar twin sample. This value agrees with those obtained by the HARPS team using RV standard stars \citep{loc15}. To properly account for the effect of the upgrade and its uncertainty, we include the RV offset as a free parameter in the model as described below.

All HARPS RVs and relevant activity indices are given in Table \ref{tbl:rvdata}.

\subsection{Analysis and results}

\begin{figure}
\resizebox{\hsize}{!}{\includegraphics{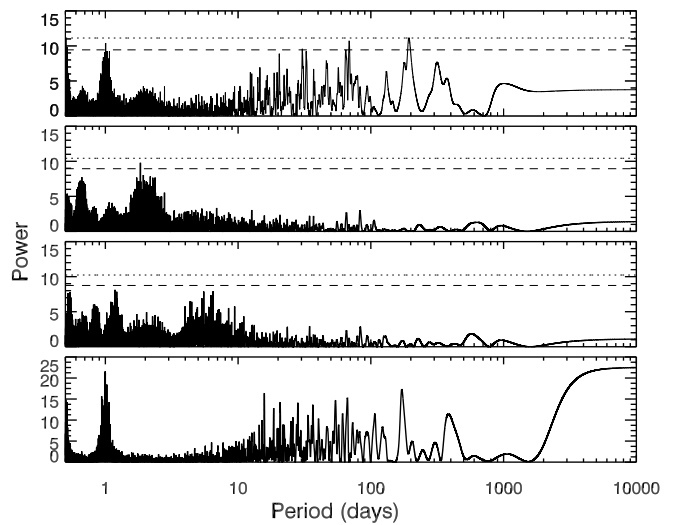}}
\caption{
Generalised Lomb Scargle (GLS) periodograms to determine the frequencies of planet candidates.  Top panel is a periodogram of the original data with instrumental upgrade offset removed.  Middle panels are periodograms of the residuals to a 194-day Keplerian fit (upper middle) and a two-planet fit (lower middle).  The horizontal lines represent false-alarm probability levels of 10\% (dashed) and 1\% (dotted), as calculated from a bootstrap Monte Carlo method.  Also included for comparison is the window function of the sampling (bottom panel).}
\label{fig:periodograms}
\end{figure}

We identified candidate planet periods using a generalised Lomb-Scargle periodogram \citep{Zechmeister2009}. 
For this initial analysis, we subtracted a 15.4 m s$^{-1}$ offset from the data taken after the HARPS upgrade. 
Leaving aside the 1.0 and 0.5-day peaks that commonly arise from a nightly sampling cadence, the strongest periodicity in the data is at 194 days (Fig. \ref{fig:periodograms}).  To calculate its false-alarm probability (FAP), or the probability that such a high-power peak could appear through random noise, we used a bootstrap Monte Carlo technique.  The data were bootstrap sampled for 5000 trials, with the power of the highest peak recorded for each trial.  The 90$^{th}$ and 99$^{th}$ percentiles of the resulting power distribution can be used as approximate 10\% and 1\% FAP thresholds.  For the 194 day peak, we find that only 0.8\% of the trials yielded a peak of equal or greater height by chance.  The analytic FAP, using the formulation of \citet{Zechmeister2009}, is even lower at 0.2\%.  
Although 194 days is somewhat close to half  a year and the associated window function peak at 171 days, the phase coverage of the signal is nonetheless sufficient to reliably fit a Keplerian. No combination of significant periodicities from the window function and the data would feasibly combine to produce an alias peak at a 194-day period \citep{daw10}.

We modelled and removed the 194 day signal using a least-squares algorithm to simultaneously fit the Keplerian orbital parameters and the RV offset introduced by the HARPS upgrade ($\delta_{inst}$). The periodogram of the residuals to this fit showed a peak at 1.84 days with a bootstrap FAP of 2\%. Adding a Keplerian with a 1.84-day period to the least-squares model fit yielded a two-planet solution which substantially reduced the residuals of the one-planet fit.
Further observations need to be gathered to better constrain the inner planet, as an alternative solution with a period of $\sim$2.3 days seems to exist.

After removing both 194-day and 1.84-day signals, no significant peaks remained in the residuals periodogram (Figure \ref{fig:periodograms}).

We ran a Markov chain Monte Carlo (MCMC) analysis using the specific implementation of \citet{bed15} to obtain final confidence intervals on the two planet candidate orbits, properly marginalising over all uncertain parameters. Our model consisted of two Keplerian signals, a constant RV offset $C$ relative to the instrumental zero point, an instrumental offset $\delta_{inst}$ for post-HARPS upgrade RVs, and a jitter term $\sigma_{J}$. The Keplerian signal was parameterised as  ($P$, $K$, $e$, $\omega$ + $M_0$, $\omega$ - $M_0$), where $P$ is the orbital period, $K$ is the RV semi-amplitude, $e$ is eccentricity, $\omega$ is the argument of periastron, and $M_0$ is the mean anomaly at a reference date set by the first RV measurement in the time series. All of the fit parameters were given uniform priors except for $P$ and $K$, which were sampled in log-space with a log-uniform prior, and $\delta_{inst}$, which was given a Gaussian prior with the mean and uncertainty as given in Sect. \ref{data}. After the MCMC run concluded, the chains were examined for non-convergence using the Gelman-Rubin statistic and any chains that had become stuck in a low-likelihood region of parameter space were removed from the posterior \citep{gel92}.

\begin{figure}
\resizebox{\hsize}{!}{\includegraphics{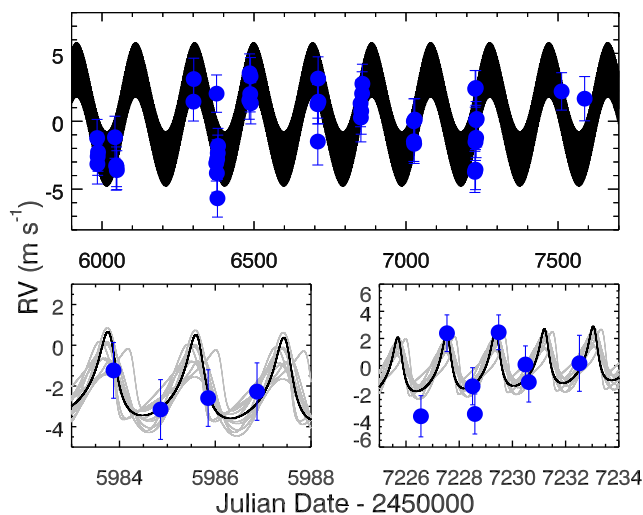}}
\caption{Full time series of RV data (blue) and two-planet best-fit model. The two lower panels show zoomed-in subsets of the time series.  The black line is the best-fit model from a least-squares fit.  The grey lines are models randomly drawn from the one-sigma range of best steps in the MCMC analysis.  As illustrated by the grey models, the eccentricity of the inner planet is not well constrained.}
\label{fig:timeseries}
\end{figure}

\begin{figure*}
\resizebox{\textwidth}{!}{\includegraphics{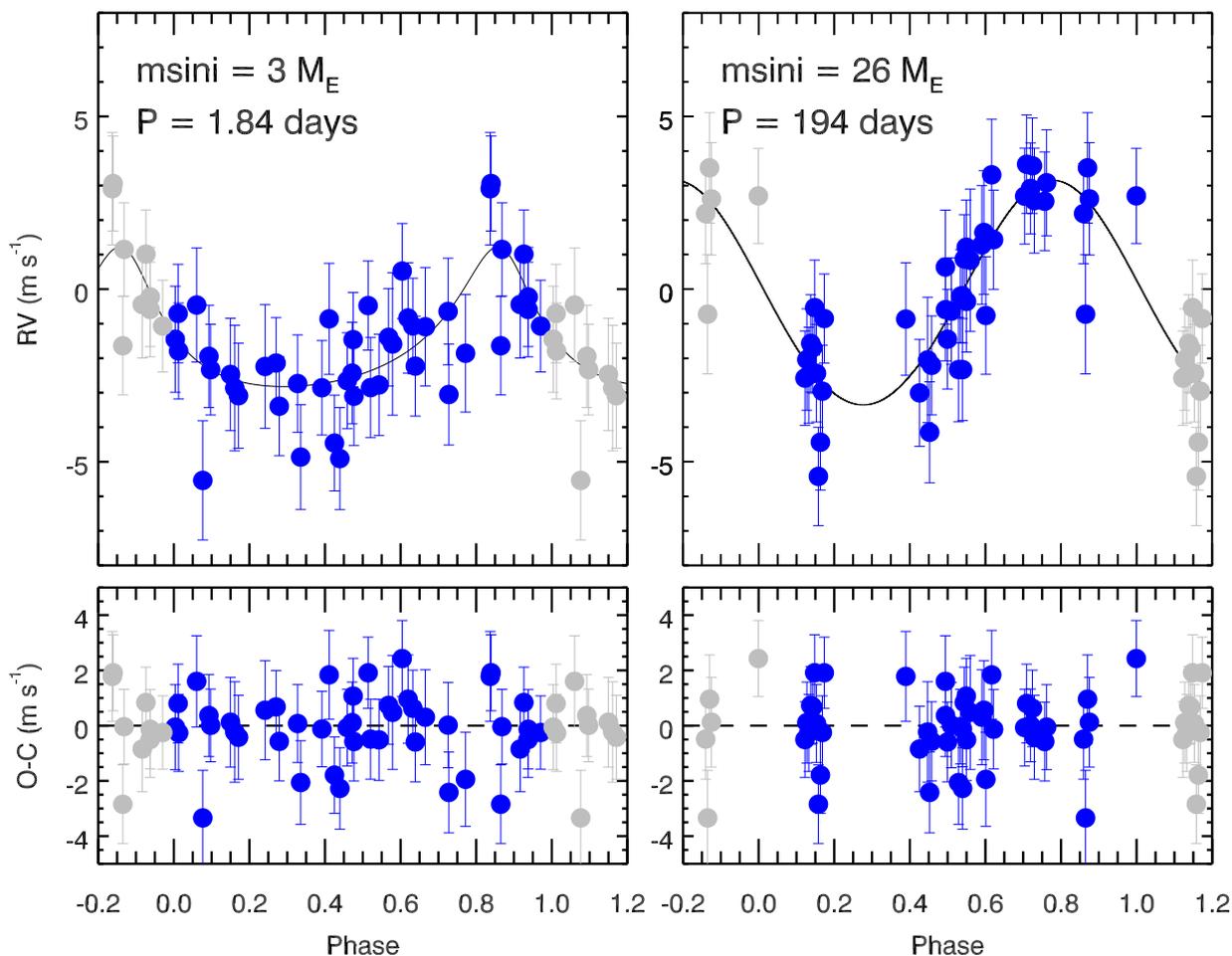}}
\caption{Phase-folded RV data for the two Keplerian signals. The upper panels show the RV data with the other planet and offsets removed. Our final MCMC solution is overplotted, with residuals in the lower panel.  Grey points are duplicated observations plotted for clarity.}
\label{fig:orbit}
\end{figure*}

The resulting constraints on the two-planet fit are presented in Table \ref{tbl:mcmc}.  Owing to strong correlations between the parameters and poorly constrained phases, taking the median values from each parameter posterior does not result in a good fit.  We quote the best-fit values from the maximum likelihood fit and MCMC median values in Table \ref{tbl:mcmc}.  The quoted errors are the one-sigma range, percentile-wise, from the MCMC posteriors.  The orbital solutions shown in Figs. \ref{fig:timeseries} and \ref{fig:orbit} use the best-fit parameters.

The final solution from the MCMC corresponds to a $2.9 \pm 0.8$ Earth mass planet with an orbital period of 1.8374 $\pm$ 0.0003 days
 and a 26 $\pm$ 4 Earth mass planet with an orbital period of 194 $\pm$ 2 days.  Both eccentricities are consistent with zero within 2$\sigma$.  
The jitter term is low (consistent with zero), indicating that HIP 68468 is a relatively quiet star without much activity.

\subsection{Model comparison}

Periodograms are a useful tool for identifying signal periods of interest, but the periodogram FAP is not the most robust tool available to assess the reality of signals found.  We ran multiple MCMCs to fit alternative models to the data.  All models include the instrumental offset factors and a jitter term, but vary in the number of Keplerian signals included.  We compared the models using the Bayesian information criterion \citep[BIC, ][]{kass1995}, calculated from the highest likelihoods achieved in the MCMCs for each model.  The results indicate that the two-planet model including a 1.84-day super-Earth has the smallest BIC and is therefore the most likely solution of those tested (see Table \ref{tbl:models}).  We also compared the best-fit solutions for each model as determined by a least-squares algorithm with similar results.

Evidence from the BIC and $\chi_{red}^2$ strongly supports the presence of the outer planet: the $\Delta$ BIC is large at 11.5, and an F-test comparing the one-planet model to the no-planet model gives a probability of about $10^{-4}$ against the planet presence.  Inclusion of the super-Earth signal improved the fit further, reducing the $\chi_{red}^2$ from 1.6 to 1.0 and bringing the RMS down from 1.7 to 1.3 m s$^{-1}$. An F-test comparison using the best-fit chi-squared statistics gives a probability of $7 \times 10^{-4}$ that the inclusion of the second planet is unwarranted.

\subsection{Stellar activity indicators}

Although HIP 68468 is a quiet star with a low log(R'$_{HK}$) value (averaging -5.05 in the HARPS spectra), we nevertheless examined the data for signs that stellar activity might be inducing a false RV signal.  No statistically significant trends were found between RV and $\log{(R'_{HK})}$ or bisector inverse span (BIS).  A marginally significant correlation was found between RV and the full width at half maximum (FWHM) of the cross-correlation function, with a p-value of 1.4\% from bootstrap resampling.  Subtracting this trend from the RV data does not have a noticeable effect on the shape of the periodogram; a peak around 190 days still stands out, although the significances of all peaks are reduced.

\begin{figure}
\resizebox{\hsize}{!}{\includegraphics{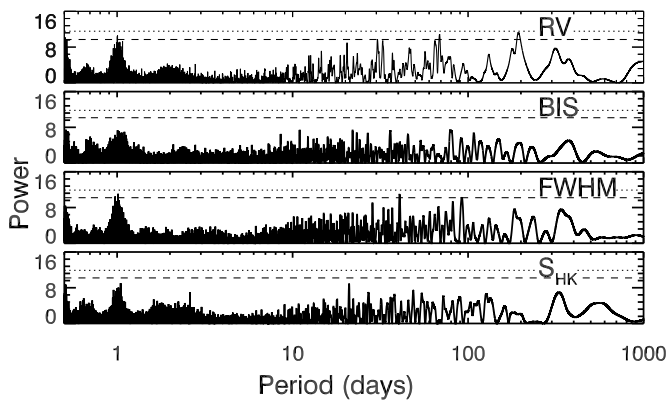}}
\caption{GLS periodograms of the RV data and various activity indicators for comparison.  Horizontal lines correspond to false-alarm probabilities of 10\% (dashed) and 1\% (dotted), as calculated from bootstrap Monte Carlo resampling.}
\label{fig:activity}
\end{figure}

The periodogram of the FWHM shows a peak at 41 days, which may be a signature of the rotation period (Fig. \ref{fig:activity}).  We were unable to fit and subtract any quasi-periodic 41-day activity signals from the data as suggested in \citet{Dumusque2011},
for example. This could be due to the limitations of the time-series sampling in resolving any evolving stellar activity. We instead treated the relationship between activity indicator and activity-induced RV shift as a linear trend with an unknown slope, which should be accurate as a first-order approximation \citep{Dumusque2014}.

We ran the MCMC algorithm with models incorporating a linear FWHM correlation in addition to the Keplerian signal(s). Because the FWHM measurement suffered the same unknown offset effect as the RVs after the HARPS upgrade, we accounted for this in a similar way by adding an additional $\delta_{inst,FWHM}$ term to the model with a strong prior based on the average FWHM offset among our constant stars of 15.6 $\pm$ 0.4 m s$^{-1}$. 

Regardless of initial conditions, the strength of the FWHM correlation term consistently went to a low value (0.13 $\pm$ 0.05).  All of the orbital parameters for the two-planet solution remained well within 1 $\sigma$ of the values given in Table \ref{tbl:mcmc}.  We therefore conclude that any stellar activity traced by the FWHM is most likely independent of the Keplerian signals in the data.  We chose not to include the FWHM correlation in the final fit presented based on the weak evidence for its addition to the model: a marginally higher BIC by 1 compared to the two-planet model, and an F-test probability of 8\% that the addition of a correlation term is unwarranted. We note, however, that the two-planet model remains the best fit regardless of whether or not an FWHM correlation is included (Table \ref{tbl:models}).

\subsection{Orbital eccentricities}

\begin{figure}
\includegraphics[scale=0.56]{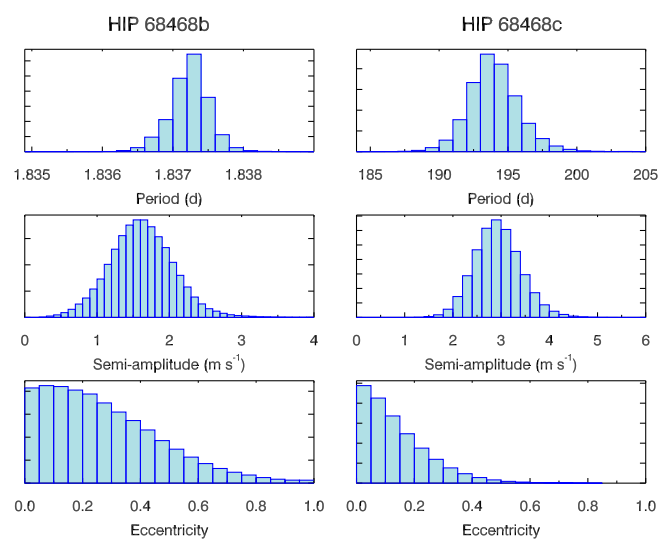}
\caption{Posterior distributions of orbital period, radial velocity semi-amplitude, and eccentricity for the two planet candidates.}
\label{fig:histograms}
\end{figure}

Both the MCMC and least-squares fitting yield non-zero eccentricities for the two Keplerian orbits.  However, as a positive definite parameter, eccentricity is prone to overestimation when an orbit is truly circular \citep{Zakamska2011}.  We present the posterior distributions of several key orbital parameters, including eccentricity, in Fig. \ref{fig:histograms}.  From the posteriors, it is clear that the mode of the eccentricity distribution \citep[recommended as the least biased estimator by][]{Zakamska2011} is within 1$\sigma$ of zero for the two planet candidates.  Nevertheless, we chose not to fix either eccentricity to zero in the MCMC analysis because the uncertainty on eccentricity should be marginalised over when making an estimate of the errors on all parameters.

The eccentricity found for the innermost planet (HIP 68468b) is remarkably high, albeit as discussed above,
within the uncertainties $e$ = 0. If the high eccentricity is confirmed with further RV observations, it
would be a short-lived phenomenon, as the orbit should be circularised on a timescale $<<$ 1 Gyr.
The exact timescale for tidal circularisation would be better defined once a more precise eccentricity is
available and when a measurement of the planet radius is made through transit observations, so that we can estimate the planet composition. 
We have recently obtained observing time with the {\em Spitzer Space Telescope} to attempt the transit detection of HIP 68468b.

\begin{savenotes}
\begin{table*}
\caption{Best-fit parameters and uncertainties for HIP68468 b and c.}
\label{tbl:mcmc}
\centering 
\begin{tabular}{llcccc} 
\hline    
\hline 
{}& &  \multicolumn{2}{c}{HIP68468 b}   & \multicolumn{2}{c}{HIP68468 c}  \\
\multicolumn{2}{c}{Parameter} & Best-Fit Value &  MCMC Posterior & Best-Fit Value & MCMC Posterior  \\
\hline
$P$ &[days] & 1.8372 & $1.8374 \pm 0.0003$ & 194 & $194 \pm 2$ \\
$K$ &[m s$^{-1}$] & 2.0 & $1.5 \pm 0.5$ & 3.3 & $2.9 \pm 0.5$ \\
$e$ & & 0.41 & $0.24^{+0.24}_{-0.17}$ & 0.04 & $0.11^{+0.14}_{-0.08}$ \\
$\omega$ + $M_0$  &[rad] & 3.0 & $2.7 \pm 1.7$ & 0.0 & $3.7^{+1.9}_{-2.5}$ \\
$\omega$ - $M_0$  &[rad] & 2.5 & $2.4^{+0.5}_{-0.4}$ & 1.0 & $1.0^{+0.2}_{-0.3}$ \\
\hline   
$m_p$ sin($i$)  & [M$_\textrm{Earth}$]  & 3.5 & $2.9 \pm 0.8$ & 30 & $26 \pm 4$ \\
$a$ \footnotemark[1] & [AU]  & 0.029743 & $0.029744 \pm 0.000003$ & 0.665 & $0.664 \pm 0.004$ \\
\hline
\hline
$C$ &[m s$^{-1}$] & 1257.2 & $1256.2^{+1.1}_{-0.8}$ & & \\ 
$\delta_{inst}$ &[m s$^{-1}$] & 15.4 & $15.5 \pm 0.2$ & & \\
$\sigma_{J}$& [m s$^{-1}$] &  & $0.4^{+0.4}_{-0.3}$ & & \\
\hline
\hline
RMS& [m s$^{-1}$] & 1.3 & & & \\
\hline
\end{tabular}
\newline
\tablefoottext{1}{Using host star mass of 1.05 M$_{\odot}$ (see text); error estimates do not include uncertainty on the stellar mass.}
\end{table*}
\end{savenotes}

\begin{table*}
\caption{Comparison of RV models.}
\label{tbl:models}
\centering 
\begin{tabular}{lccccc} 
\hline    
\hline 
{ } & BIC & $\chi^2_{red}$ & RMS [m s$^{-1}$] \\
\hline
No planets & 217.8 & 2.9 & 2.4 \\
No planets + FWHM correlation & 222.8 & 2.6 & 2.3 \\
194-d Keplerian & 206.3 & 1.6 & 1.7 \\
194-d Keplerian + FWHM correlation & 211.0 & 1.6 & 1.7 \\
194-d \& 1.84-d Keplerians & 199.0 & 1.0 & 1.3 \\
194-d \& 1.84-d Keplerians + FWHM correlation & 200.2 & 0.9 & 1.2 \\
\hline
\end{tabular}
\end{table*}

\section{Fundamental parameters and abundance analysis}

\subsection{Data and measurements}
The spectra of HIP 68468 and the Sun (reflected light from the asteroid Vesta) were taken using the MIKE spectrograph \citep{ber03} at the 6.5m Clay Magellan telescope.
The data were obtained with R = 65 000 in the red side (500 - 1000 nm) and R = 83 000 in the blue side (320 - 500 nm).
The spectra have a signal-to-noise ratio (S/N) $\approx$ 400 per pixel at 600nm. The data were reduced using the CarnegiePython MIKE pipeline\footnote{http://code.obs.carnegiescience.edu/mike}, and further processing (Doppler correction and continuum normalisation) was performed as described in \cite{ram14}.
The MIKE data were employed to perform the equivalent width (EW) measurements used to determine stellar parameters and chemical abundances,
and for spectral synthesis to determine $v$ sin $i$ and the lithium content.

The HARPS spectra described in Sect. 2.1 have a more limited wavelength coverage (380-690 nm) than the MIKE
spectra, but they have a higher resolution (R = 120 000), hence they were used to verify through spectral synthesis both
$v$ sin $i$ and the Li abundance. The combined HARPS spectrum of HIP 68468 has S/N $\sim$750 at 600nm.

The EW were measured comparing line-by-line the spectrum of HIP 68468 to the Sun in a 6 \AA\ window, 
to help determine the continuum and to choose the part of the line profile that was used to fit a Gaussian profile.
After the first set of measurements was obtained, 2-$\sigma$ outliers were identified from the differential analysis
and their EW were verified. The remeasured EW were kept, unless there was a problem,
such as a contamination by a telluric line or a defect in one of the spectra.

\subsection{Stellar parameters}

The stellar parameters for our solar twin planet search sample were published in \cite{ram14}.
Only three of the sample stars had revised stellar parameters
based on the [Y/Mg] ratio, which can be used as a proxy for stellar ages \citep{nis15}.
The comparison between our isochrone ages \citep{ram14} and the [Y/Mg] ratio \citep{tuc16}
shows that [Y/Mg] ages can be determined to better than 0.9 Gyr.
As the isochrone age of HIP 68468 determined in \cite{ram14} was off by 1.2 Gyr relative to the 
age obtained from the [Y/Mg] ratio, and because the age of this star plays an important role 
in our interpretation,  a reanalysis of HIP 68468 was performed as described below.

We adopted the same line list as in \cite{ram14}. 
The EW measurements were used to obtain elemental abundances with the {\em abfind} driver of MOOG \citep{sne73}, 
using Kurucz model atmospheres \citep{cas04}.  Then, line-by-line differential abundances were obtained.

The stellar parameters (\tsin, log $g$, [Fe/H], $v_t$) of HIP 68468 were determined by imposing a 
differential spectroscopic equilibrium of iron lines relative to the Sun \citep[e.g.][]{mel14a}, 
 using as initial parameters the values given in \cite{ram14}. The solar values were kept fixed at
 (\tsin, log $g$, $v_t$) = (5777 K, 4.44 dex, 1.0 km s$^{-1}$). 
 Our solution for HIP 68468 gives
 consistent differential abundances for neutral and ionised species of different elements, and also for atomic and
 molecular lines, as discussed in more detail in the next section.
 
 The errors in the stellar parameters were obtained from the uncertainties in the slope of the fits of the differential abundances versus
 excitation potential and reduced EW, and the errors in the differential ionisation equilibrium (based on the
 observational uncertainties in the differential abundances of FeI and FeII), and
 including also the degeneracy among the stellar parameters.
 
 The resulting stellar parameters are \teff = 5857$\pm$8 K, log $g$ =  4.32$\pm$0.02, [Fe/H] = 0.065$\pm$0.007, $v_t$ = 1.14$\pm$0.01 km s$^{-1}$,
 which agree well with the previous estimate (5845$\pm$6 K, 4.37$\pm$0.02 dex, 0.054$\pm$0.005 dex, 1.13$\pm$0.01 km s$^{-1}$) by \cite{ram14}.
 The most important difference is seen in the log $g$ value, which is reduced by 0.05 dex. 
 This has a non-negligible effect on the derived age of the star (0.7 GYr, see below). HIP68468 is one of the most distant stars in our 
 solar twin planet search sample, which prevents us from using its Hipparcos parallax to better constrain this important parameter. The Hipparcos parallax of 
 HIP68468 has an error greater than 10\% and the trigonometric 
 log $g$ value that we derive using that parallax is fully consistent with both the old and new spectroscopic parallaxes within the errors.
 
\begin{figure}
\resizebox{\hsize}{!}{\includegraphics{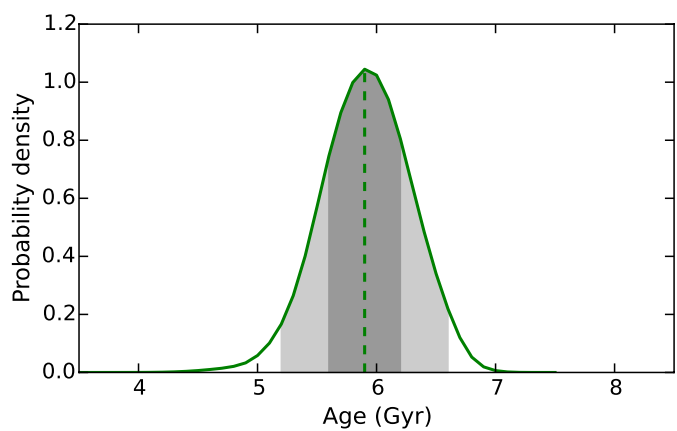}}
\caption{
Age probability distribution of HIP 68468. The dark and light areas correspond to 1$\sigma$ and 2$\sigma$
confidence.
}
\label{age}
\end{figure}

The mass and age of HIP 68468 were determined using Y$^2$ isochrones \citep{dem04}, as described in \cite{mel12} and \cite{ram13}. 
With our stellar parameters and their uncertainties, we used probability distribution functions to infer a mass M = 1.05$\pm$0.01 M$_\odot$
and age of 5.9$\pm$0.4 Gyr (Fig. \ref{age}), which are similar to those reported by \cite{ram14}, 1.04$\pm$0.01 M$_\odot$ and 5.2$^{+0.4}_{-0.5}$ Gyr.

Recently, \cite{nis15} has shown a tight correlation between [Y/Mg] and age that is corroborated by our own
study \citep{tuc16}. HIP 68468 has [Y/Mg] = $-$0.087$\pm$0.017 dex, implying in an age of 6.4$\pm$0.8 Gyr
\citep[using the relation by][]{tuc16}. This age agrees well with our isochrone age (5.9$\pm$0.4 Gyr). 
Thus, both isochrones and [Y/Mg] suggest
that HIP 68468 has an age of about 6 Gyr.

The post-solar age of HIP 68468 is further supported by its low activity level. The HARPS spectra give an activity index log($R'_{HK}$) = -5.05 $\pm$ 0.02, where the error comes from the standard deviation across all spectra in the time series.  Using the relationship derived by \citet{Mamajek2008} and a 
stellar (B-V) of 0.68 \citep{Tycho2000}, this translates into an expected age of 7.6 $\pm$ 0.4 Gyr, but we note that the activity-age relation is not
well defined above 3 Gyr \citep{ram14}, hence the above error bar is underestimated.

The projected rotational velocity of HIP 68468 was estimated through spectral synthesis by \cite{leo16}, 
taking into account both the instrumental ($\Delta \lambda = \lambda/R$) and macroturbulent broadening \citep[using a 
radial-tangential macroturbulence profile;][]{gra05}. 

We note that if the macroturbulence ($V_{\rm macro}$) is
estimated from $V_{\rm macro}$-\teff relations \citep[e.g.,][]{mel12},
then HIP 68468 seems to be rotating faster than it should for its age,
when compared to other solar twins analysed with the same method.
However, even for main-sequence stars the luminosity effect on macroturbulence
is not negligible \citep{doy14}. 
\cite{leo16} provided a new macroturbulence calibration that takes the \teff  and log $g$ dependence into account.
This gives a projected rotation velocity for HIP 68468 of $v$ sin $i$ = 1.92$\pm$0.13 km s$^{-1}$
\citep{leo16}, which agrees within the errors with the rotation velocity predicted  by the v$_{\rm rot}$-age relation by \cite{leo16}
at HIP 68468's age (5.9 Gyr), v$_{\rm rot}$ = 1.86 km s$^{-1}$.

\subsection{Stellar abundances}

The chemical composition was determined using the line list by \cite{mel14a} for atomic lines.
To verify the carbon abundances obtained from permitted atomic lines, we
also obtained differential abundances using the molecules CH and C$_2$, adopting the
molecular data given in \cite{asp05}.

For the elements V, Mn, Co and Cu, we took hyperfine structure (HFS) into account.
The differential HFS corrections are small because of the similarity between HIP 68468 and the Sun.
Differential NLTE corrections in solar twins are also usually small, as shown in 
our previous works \citep{mel12,mel14a,mon13}. 
Still, we performed NLTE corrections for Li \citep{lin09}, O \citep{ram07}, Na \citep{lin11}, Mg, 
Ca (Lind et al., in prep., see appendix A), and Cu  \citep{shi14,yan15}. 
The differential NLTE corrections are small for most elements, amounting to 
$-0.004$, $-0.006$, $-0.004$ and 0.000 dex for Na, Mg, Ca, and Cu, respectively.
Only for the oxygen triplet was the differential NLTE effect strong, amounting to $-0.028$ dex.

\begin{figure}
\resizebox{\hsize}{!}{\includegraphics{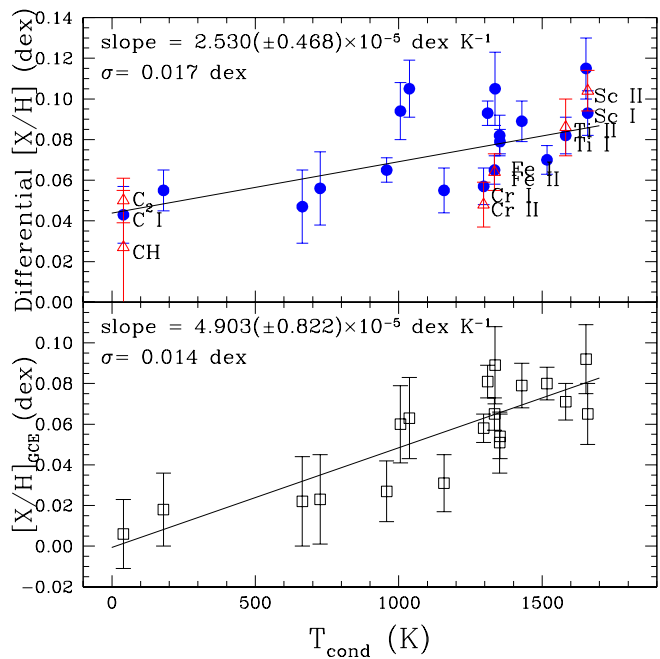}}
\caption{
{\it Upper panel:}. Observed abundance pattern of HIP 68468 versus condensation temperature
before GCE corrections. 
The solid line is a fit taking into account the total error bars.
It has a significance of 5.4-$\sigma$, and the element-to-element scatter from the fit is 0.017 dex,
which is larger than the average error bar (0.012 dex).
{\it Lower panel:}. Abundance pattern after GCE corrections to
the solar age. The error bars also include the error of the GCE corrections,
and has an average of 0.014 dex. The correlation with condensation temperature is now
stronger and more significant (6.0-$\sigma$), and the element-to-element scatter about
the fit (0.014 dex) is in perfect agreement with the average error bar.
}
\label{tcond}
\end{figure}

The final differential abundances, after taking into account HFS and NLTE effects, are listed in Table \ref{abund},
where LTE abundances are also shown for completeness.
These abundances are plotted as a function of condensation temperature in Fig. \ref{tcond}. The upper panel) shows
the excellent agreement between neutral and ionised species (shown by triangles) for the elements Cr, Fe (forced by the analysis), Ti, and Sc.
The carbon abundance obtained from neutral carbon atomic lines
and those obtained from CH and C$_2$ lines (both shown by triangles in Fig. \ref{tcond}) also agree well. 
This consistency supports our determination of stellar parameters,
as lines from different species have different dependence on the stellar parameters.

The abundance errors are reported in Table \ref{abund} and are based on the measurement errors
(standard error) and systematic errors due to the uncertainties in the stellar parameters. The total errors are on average 0.012 dex.
This is similar to the 0.01 dex errors achieved with
high-quality spectra of stellar twins \citep[e.g.][]{mel12,mon13,bed14,ram15,tes16}.

\begin{figure}
\resizebox{\hsize}{!}{\includegraphics{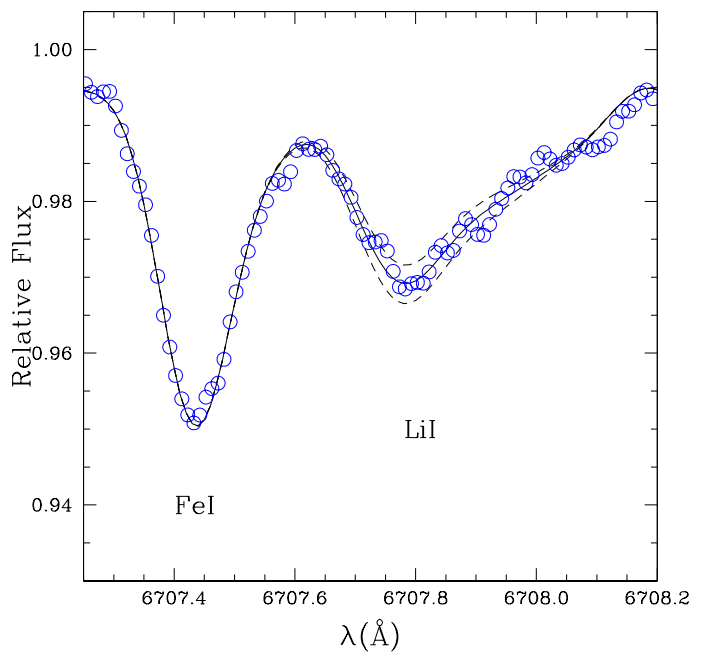}}
\caption{
Lithium feature in the HARPS spectrum of HIP 68468 (open circles) and synthetic spectrum 
with an LTE lithium abundance of 1.47 dex (solid line), corresponding to an NLTE abundance of A(Li) = 1.51 dex.
A variation of  $\pm$0.04 dex in the Li abundance is shown by dashed lines.
The variations seen in the line profile are mostly consistent with the noise in this region (S/N $\sim$ 500),
but could also be due to distinct convective line shifts between different species.
}
\label{lifeature}
\end{figure}

The lithium feature is clearly visible and deeper than in the Sun,
so that a reliable lithium abundance can be estimated.
The Li content was obtained using spectral synthesis in LTE, including blends by
atomic and molecular (CN, C$_2$) lines, employing the line list of \cite{mel12}.
Using the MIKE spectrum, we obtain an LTE lithium abundance of A(Li) = 1.49$\pm$0.05,
and our HARPS spectrum results in A(Li) = 1.47$\pm$0.04 (Fig. \ref{lifeature}).
We adopted an LTE abundance of A(Li) = 1.48$\pm$0.03 (weighted average), and using the NLTE corrections 
mentioned above \citep{lin09}, A(Li) = 1.52$\pm$0.03 was obtained.

\begin{figure}
\resizebox{\hsize}{!}{\includegraphics{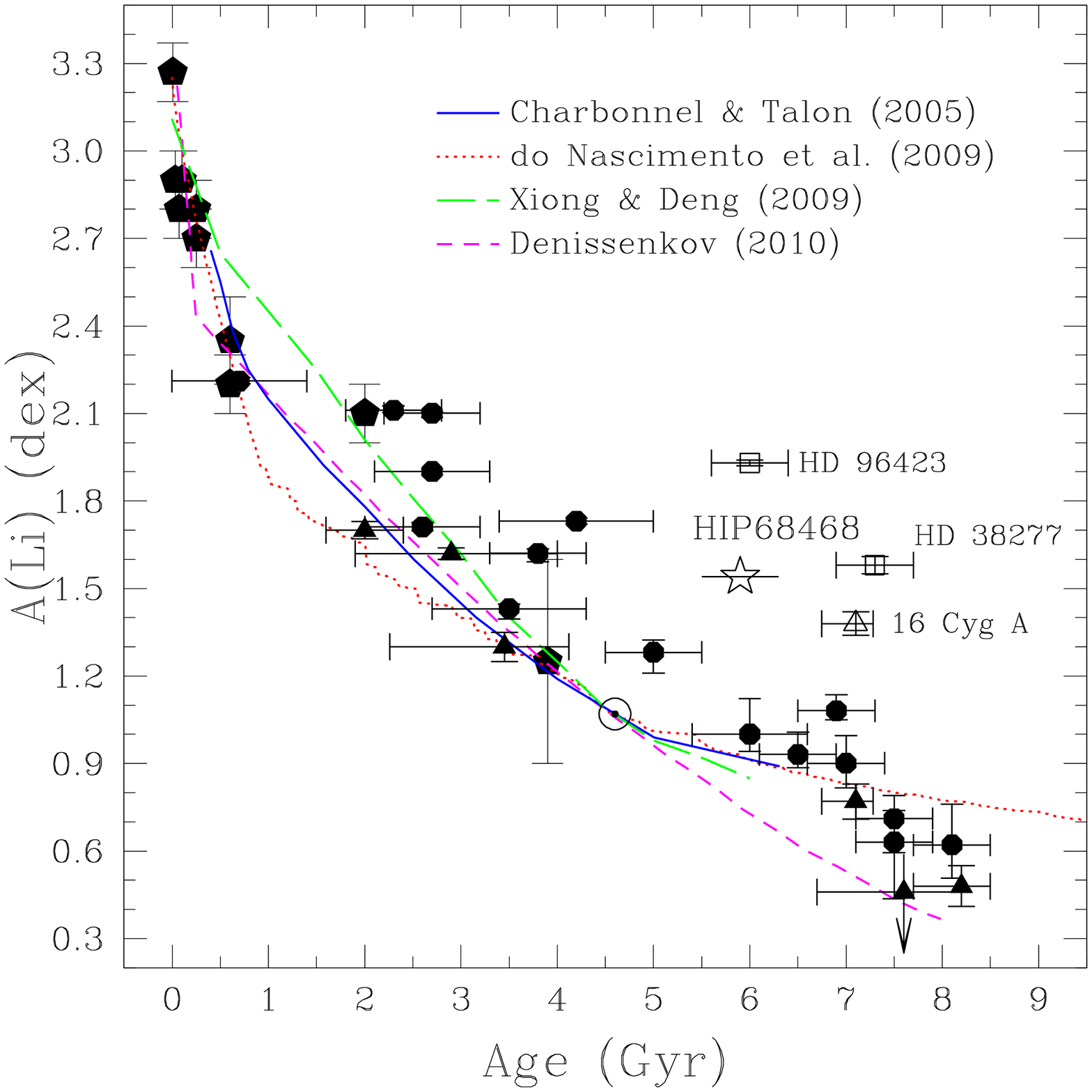}}
\caption{
Lithium abundances versus age. The Li abundance of HIP 68468 was determined in this work.
For the field solar twins (circles, triangles, squares) we used the values given in our 
previous works \citep{car16,jhon16a,mel12,mel14a,mel14b,mon13,ram11},
and for the solar twins in open clusters (pentagons) we adopted the values given in \cite{bau10} and
\cite{cas11}, as described in the text.
We also show non-standard models of lithium depletion \citep{ct05,don09,xd09,den10},
normalised to the Sun.
}
\label{lithium}
\end{figure}

The Li abundance in HIP 68468 (1.52$\pm$0.03 dex) is much higher (four times; 0.6 dex) 
than expected for its age \citep{mon13,mel14b,car16}. This is shown in Fig. \ref{lithium}, where
HIP 68468 is compared with solar twins analysed in our previous works \citep{ram11,mel12,mel14a,mel14b,mon13,car16,jhon16a}
and to solar twins in open clusters using data by \cite{bau10}, except for the cluster M67,
where the Li abundances are from the updated results using spectrum synthesis by \cite{cas11}.
The solar twins in M67 were selected from their list of solar analogs (Table 1), considering \teff = \teff$_\odot \pm$100K and $L$ = 1.0$\pm$0.1 $L_\odot$.
The mean LTE Li abundance in M67 solar twins is 1.25 ($\sigma$ = 0.35 dex), with the scatter being due partly to
the uncertain values (some are only upper limits) given in \cite{cas11}. Nevertheless, the
mean Li abundance is similar to the LTE Li abundance (1.26 dex) 
of the only solar twin (M67-1194) analysed at high precision in this cluster \citep{one11}. 
Although there might be some slight variations in Li as a result
of slightly different masses and metallicities,
contributing thus to the observed scatter in the Li-age correlation,
the variations for our small range in mass and metallicity are small \citep{bau10,car16}, 
hence our comparison between HIP 68468 and other solar twins is valid.

The decay of lithium with age is supported by stellar evolution models 
including transport processes beyond merely the mixing length convection
\citep{ct05,don09,xd09,den10},
as shown in Fig. \ref{lithium}. Lithium is depleted and not produced in stars like the Sun, therefore
the enhanced lithium abundance of HIP 68468 is probably due to external pollution.
\cite{car16} have also recently identified two stars with enhanced lithium abundances
(open squares in Fig. \ref{lithium}) and
suggested that they may have been polluted in lithium by planet ingestion,
which might also be the case of 16 Cyg A (open triangle in Fig. \ref{lithium}).

There are four stars with enhanced lithium abundances in the sample of 27 thin-disc solar twins 
shown in Fig. \ref{lithium} (HIP 68468, HD 38277, HD 96423, 16 Cyg A).  
Adopting a binomial distribution \citep[see e.g. Sect. 4.4 in][]{ram12},
the above numbers suggest that 15$\pm$2\% of Sun-like stars may have visible signatures of
planet ingestion.

\begin{table*}
\caption{Differential abundances of HIP 68468 relative to the Sun and errors.}
\label{abund}
\renewcommand{\footnoterule}{}  
\begin{tabular}{lrrrrrrrrrr} 
\hline    
\hline 
{Element}& [X/H]   & $\Delta$ \teff & $\Delta$log $g$ & $\Delta v_t$ & $\Delta$[Fe/H] & param\tablefootmark{a} & obs\tablefootmark{b} & total\tablefootmark{c} \\
{}       &       &   +8K           &  +0.02dex      & +0.01 km s$^{-1}$  & +0.01 dex   &  &  &  \\
{}       & (dex) & (dex) & (dex)         & (dex)           & (dex)       & (dex)        & (dex) & (dex) \\
\hline
Li   & 1.52\tablefootmark{d} & 0.01 &  0.00 &  0.00 &  0.00 & 0.01 & 0.03 & 0.03\\
C   & 0.043 & -0.005 &  0.005 &  0.000 &  0.000 & 0.007 & 0.012 & 0.014\\
CH  & 0.027 &  0.006 & -0.001 &  0.000 &  0.007 & 0.009 & 0.026 & 0.028\\
C$_2$  & 0.050 &  0.007 &  0.001 &  0.000 &  0.007 & 0.010 & 0.005 & 0.011\\
O   & 0.055 & -0.008 &  0.003 & -0.001 &  0.002 & 0.009 & 0.005 & 0.010\\
Na  & 0.065 &  0.003 & -0.001 &  0.000 &  0.000 & 0.003 & 0.002 & 0.006\\
Mg  & 0.105 &  0.006 & -0.002 & -0.003 &  0.000 & 0.007 & 0.017 & 0.018\\
Al  & 0.115 &  0.003 & -0.002 & -0.001 &  0.000 & 0.004 & 0.015 & 0.015\\
Si  & 0.093 &  0.001 &  0.001 & -0.001 &  0.001 & 0.002 & 0.006 & 0.006\\
S   & 0.047 & -0.004 &  0.005 &  0.000 &  0.001 & 0.006 & 0.017 & 0.018\\
K   & 0.094 &  0.006 & -0.008 & -0.002 &  0.001 & 0.010 & 0.009 & 0.014\\
Ca  & 0.070 &  0.005 & -0.003 & -0.002 &  0.000 & 0.006 & 0.003 & 0.007\\
Sc I & 0.093 &  0.006 &  0.000 &  0.000 & -0.001 & 0.006 & 0.009 & 0.011\\
Sc II & 0.104 &  0.000 &  0.007 & -0.002 &  0.003 & 0.008 & 0.006 & 0.010\\
Ti I & 0.082 &  0.007 &  0.000 & -0.002 & -0.001 & 0.007 & 0.006 & 0.009\\
Ti II & 0.086 &  0.000 &  0.007 & -0.002 &  0.003 & 0.008 & 0.011 & 0.014\\
V   & 0.089 &  0.008 &  0.000 & -0.001 & -0.001 & 0.008 & 0.005 & 0.010\\
Cr I & 0.057 &  0.006 & -0.001 & -0.002 &  0.000 & 0.006 & 0.006 & 0.009\\
Cr II & 0.048 & -0.003 &  0.007 & -0.002 &  0.002 & 0.008 & 0.008 & 0.011\\
Mn  & 0.055 &  0.006 & -0.002 & -0.003 &  0.000 & 0.007 & 0.008 & 0.011\\
Fe I & 0.065 &  0.006 & -0.001 & -0.002 &  0.000 & 0.006 & 0.002 & 0.007\\
Fe II & 0.064 & -0.002 &  0.007 & -0.003 &  0.003 & 0.008 & 0.004 & 0.009\\
Co  & 0.082 &  0.006 &  0.002 &  0.000 &  0.000 & 0.006 & 0.008 & 0.010\\
Ni  & 0.079 &  0.005 &  0.000 & -0.002 &  0.000 & 0.005 & 0.003 & 0.006\\
Cu  & 0.105 &  0.004 &  0.001 & -0.001 &  0.001 & 0.004 & 0.013 & 0.014\\
Zn  & 0.056 &  0.000 &  0.001 & -0.003 &  0.002 & 0.004 & 0.018 & 0.018\\
\hline       
\end{tabular}
\tablefoot{Abundances of V, Mn, Co, and Cu accounted for HFS. NLTE effects were considered for Li, O, Na, Mg, Ca, and Cu;
LTE abundances for these elements are 1.48, 0.083, 0.069, 0.099, 0.075, 0.105 dex, respectively}
\tablefoottext{a}{Adding errors in stellar parameters}
\tablefoottext{b}{Observational errors}
\tablefoottext{c}{Total error (stellar parameters and observational)}
\tablefoottext{d}{A(Li) is reported rather than [Li/H].}
\end{table*}

\section{Discussion}

The presence of a giant planet and a small super-Earth in close-in orbits gives us the opportunity 
to study how this configuration might be related to the chemical abundance pattern in HIP 68468.

In Fig. \ref{tcond} (upper panel) we show that the abundance pattern of HIP 68468 (relative to the Sun) has 
a correlation with  condensation temperature ($T_{\rm cond}$). The fit of [X/H] vs. $T_{\rm cond}$ is
represented by a solid line. This  correlation is
well defined, with a significance higher than 5-$\sigma$. The element-to-element scatter about the fit
is 0.017 dex, which is higher than the average abundance error of 0.012 dex, but this is likely due
to the scatter introduced by galactic chemical evolution (GCE), as described in recent works \citep{nis15,spi16}.

Employing the relations between age and stellar abundances obtained by \cite{spi16} using solar twins,
we can correct for GCE effects, which enables a more proper comparison to the Sun, which is 
$\Delta_{\rm age}$ = 1.3 Gyr younger than HIP 68468.  We subtracted the GCE effects corresponding 
to this age interval, resulting in abundance ratios corrected to the Sun's age,

\begin{equation}
[X/H]_{\rm GCE} = [X/H] - slope * \Delta_{\rm age},
\end{equation}

\noindent where the {\it \textup{slope}} of the GCE corrections for the different chemical elements is taken from Table 3 of \cite{spi16}. 

The corrected abundance ratios are shown in the bottom panel of Fig. \ref{tcond}. In this plot
the elements with different species were weight averaged, using as weights the inverse square of their error bars.
The error bars now also include the error on the GCE corrections, which are due to the error on the slope of
the abundance variations with age \citep{spi16} and also considering the error on the age of HIP 68468.
As a result, the typical error bar increased from 0.012 dex to 0.014 dex. The [X/H]$_{\rm GCE}$ ratios now show
a higher refractory enhancement and a stronger correlation with $T_{\rm cond}$, at the level of 6-$\sigma$.
The scatter about the fit is now 0.014 dex, identical to the typical error bar in the [X/H]$_{\rm GCE}$ ratios.

The close-in giant planet suggests migration from an outer to the inner ($\leq$ 1 AU) region.
This migration could have driven inner planets towards its host star, leading to planet engulfment events
and to changes in the abundance pattern of the convection zone \citep{san02}. The
super-Earth that we seem to have detected may also follow this fate, as it is located at only 0.03 AU from HIP 68468.
If HIP 68468b survives, it surely would be destroyed when HIP 68468 evolves from the main sequence.

Past planet accretion effects could be reflected in enhanced abundances of the refractory elements, 
and also an increase in the lithium abundance. As lithium is affected by stellar depletion by more than a factor of 100
at the solar age \citep[e.g.][]{asp09,mon13}, it is relatively easy to increase the low photospheric Li content by planet accretion.
As shown by \cite{san02}, in a planet migration scenario and further accretion to its host star, 
the planet material would be mixed in the stellar convection zone and would modify the surface abundances, in particular for Li
\citep[see also][]{tog16,mur02,mon02}. 

\begin{figure}
\resizebox{\hsize}{!}{\includegraphics{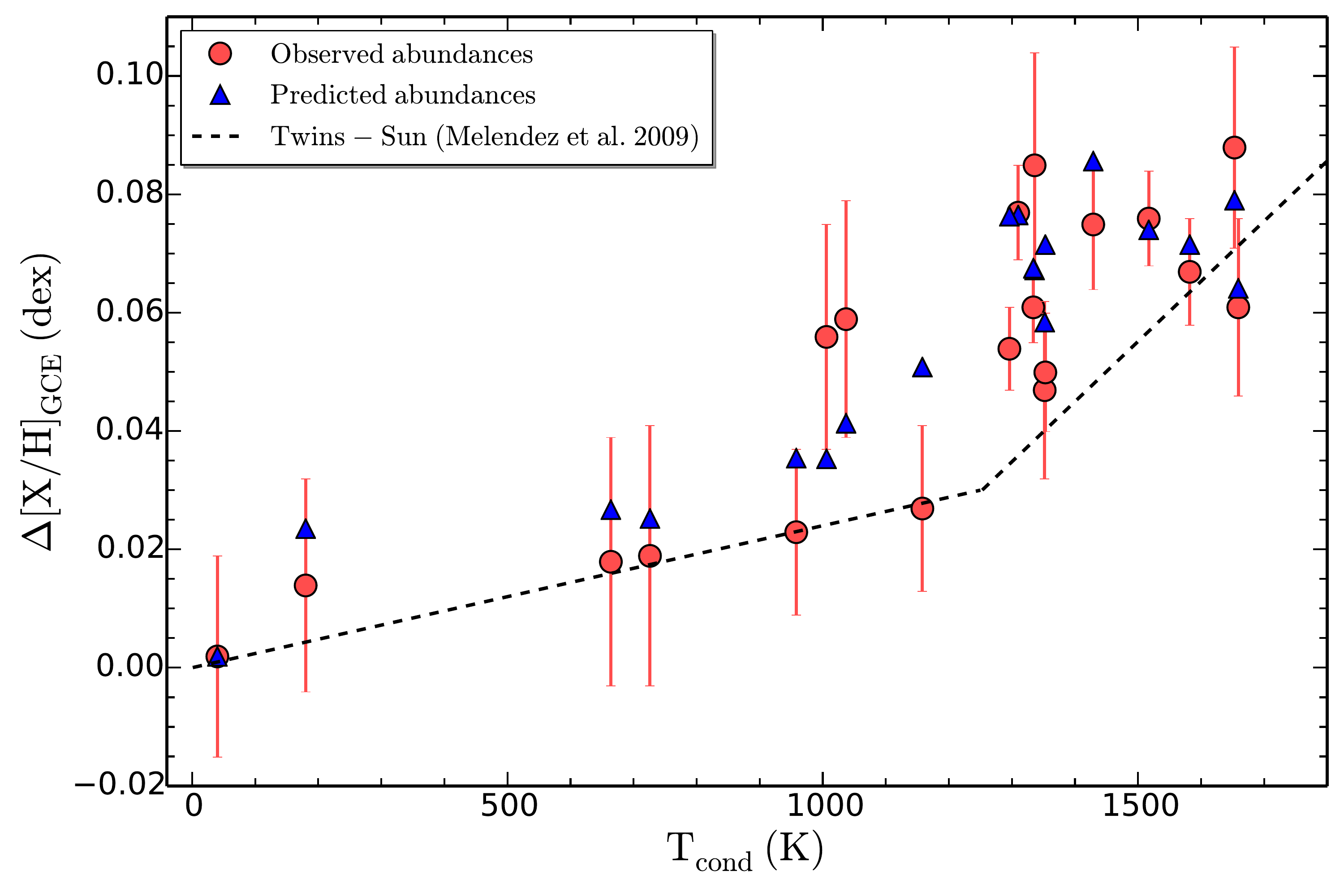}}
\caption{
Observed abundance ratios [X/H]$_{\rm GCE}$ in HIP 68468  
as a function of condensation temperature (red circles).
The abundances are corrected for galactic chemical evolution to the Sun's age. 
Planet engulfment of 6 Earth masses of rocky material (blue triangles) can explain the enhancement of 
refractory elements.
The mean abundance trend of 11 solar twins relative to the Sun \citep{mel09}
is shown by the dashed line.
}
\label{gce}
\end{figure}

Using the abundance pattern of the Earth and meteorites,
we can estimate the rocky mass needed to increase the refractory elements to the
levels observed in HIP 68468, following the procedure outlined in \cite{cha10} and
\cite{jhon16b}. We estimated a convective mass of 0.018 M$_\odot$,
using the tracks by \cite{sie00}.
In Fig. \ref{gce}, we show the [X/H]$_{\rm GCE}$ ratios by red circles and 
the effect due to the accretion of 6 Earth masses of rocky material into the convection zone
of HIP 68468 by blue triangles. The best fit is provided with a mix
of two and four Earth masses of meteoritic-like and Earth-like material, respectively.

The same amount of rocky material (6 Earth masses) is more than enough to explain its
Li enhancement. The net lithium enhancement is hard to predict because it
depends on {\it (i)} the initial stellar lithium content at the time of accretion, 
{\it (ii)} the stellar Li depletion since the planet accretion, 
{\it (iii)} the efficiency of thermohaline instabilities for
depletion of Li that is due to the planet engulfment \citep{the12}.
For example, if there is no extra depletion that is due to thermohaline mixing, 
then we find that the planet accretion event could have occurred when the
star had A(Li) $\sim$ 2 dex, corresponding to an age of about 1 Gyr.
Assuming that thermohaline mixing depletes lithium by 0.5 dex or 1.0 dex,
then the initial Li at the time of planet accretion was either 1.45 dex or 0.9 dex,
corresponding to ages of about 3 or 6 Gyr, respectively.
This means that if thermohaline mixing is very efficient, depleting Li in about 1 dex,
then the planet accretion event occurred recently.

If accretion of planetary material occurred in recent times, 
then debris may remain around the star. 
The low log(R'$_{HK}$) activity index of HIP 68468, while not obviously anomalous for its age group, might in part be due to absorption by dust in the chromospheric Ca II lines similar to the effect seen in WASP-12 \citep{Haswell2012, Fossati2013}. Further observations including photometric monitoring for anomalously shaped transits of dust clouds would test this hypothesis.
Although the activity index might be affected by the dust, the stellar activity as traced by the RVs would not. 
The low RV jitter is therefore consistent with the old age and low log(R'$_{HK}$).

\cite{mel09} and \cite{ram09} argued that the Sun is deficient in refractory elements relative
to the majority solar twins, probably because of the formation of terrestrial planets in the solar system.
Most solar twins are therefore enhanced in refractory elements relative to the Sun, such as the solar twin HIP 68468.
However, it must be stressed that the refractory enhancement of most solar twins 
may be primordial, unlike the case of HIP 68468, where the refractory enrichment seems to be due
to planet accretion at least 1 Gyr after the star was born. This distinction can be made thanks to the enhanced lithium in HIP 68468,
in contrast to most solar twins, which follow a well-defined Li-age correlation.

\section{Concluding remarks}

We have discovered two strong planet candidates, a close-in (0.66 AU) planet more massive than Neptune and a close-in (0.03 AU) super-Earth,
around the solar twin HIP 68468.
We have recently been granted {\em Spitzer} time to search for the transit of this super-Earth.  

The host star has enhanced abundances of refractory elements relative to the Sun.
After corrections for GCE to bring the [X/H] abundance ratios to the solar age, the correlation 
with condensation temperature is even stronger, and the element-to-element scatter about the fit 
with $T_{\rm cond}$ is reduced.
The lithium content of HIP 68468 is four times higher than expected for its age. 
The enhancement of Li and of the refractories can be reproduced by engulfment of
a rocky planet with 6 Earth masses.

The planet configuration of HIP 68468 suggests planet migration, which might
have resulted in planet accretion events that left signatures on the chemical composition 
of the host star, as observed for HIP 68468, and in agreement with the predictions by \cite{san02}.

The recent discoveries of stars showing chemical anomalies possibly related to planet engulfment episodes 
\citep{ash05,spi15} and the exciting case of HIP 68468 open the possibility to extend the search to other 
objects with such distinctive chemical patterns. Identifying a population of such objects will provide important 
indications of dynamical interactions of planets and the mechanisms that can drive the evolution of systems similar to our own.

While the detection of the two planets is formally statistically
significant (see Sect. 2.3), secure detection of small planets with the
radial velocity technique has typically been based on a substantially
larger number of measurements given the challenges of stellar activity and
reliable period determination with sparse sampling \citep[e.g.][]{Pepe2011}. The intriguing evidence
of planet accretion in the form of the enhanced refractory element and
lithium abundances warrants further observations to verify the existence
of the planets that are indicated by our data and to better constrain the nature
of the planetary system around this unique star.

\begin{acknowledgements}
 We would like to thank the many scientists and engineers who made HARPS possible. 
 Thanks also to Carole Haswell and Dan Staab for useful discussion. 
 J.M. acknowledges support from FAPESP (2012/24392- 2) and CNPq (Bolsa de Produtividade). 
 M.B. is supported by the National Science Foundation (NSF) Graduate Research Fellowships Program (grant no. DGE-1144082). J.B. and M.B. acknowledge support for this work from the NSF (grant no. AST-1313119). J.B. is also supported by the Alfred P. Sloan Foundation and the David and Lucile Packard Foundation.
M.A. acknowledges support from the Australian Research Council (grants FL110100012 and DP120100991).
 
\end{acknowledgements}

\newpage

\begin{table*}
\caption{HARPS measured radial velocities and activity indices for HIP 68468.}
\label{tbl:rvdata}
\centering 
\begin{tabular}{lccccc} 
\hline    
\hline 
{Julian Date} & RV (km s$^{-1}$) & $\sigma_{RV}$ (km s$^{-1}$) & $S_{HK}$ & $\sigma_{S_{HK}}$ & FWHM (km s$^{-1}$)  \\
\hline
    2455983.87576651&      1.2544&  0.0009&  0.1514&  0.0006&      7.3631 \\
    2455984.85786337&      1.2525&  0.0011&  0.1538&  0.0008&      7.3684 \\
    2455985.85166182&      1.2531&  0.0010&  0.1504&  0.0007&      7.3638 \\
    2455986.87231354&      1.2534&  0.0010&  0.1502&  0.0007&      7.3674 \\
    2456042.62854487&      1.2545&  0.0012&  0.1526&  0.0009&      7.3742 \\
    2456046.75246848&      1.2524&  0.0015&  0.1491&  0.0011&      7.3754 \\
    2456047.79316449&      1.2523&  0.0011&  0.1547&  0.0008&      7.3708 \\
    2456048.80623149&      1.2520&  0.0010&  0.1508&  0.0008&      7.3782 \\
    2456300.86637575&      1.2571&  0.0010&  0.1527&  0.0008&      7.3743 \\
    2456301.83524661&      1.2588&  0.0012&  0.1501&  0.0009&      7.3758 \\
    2456375.81375898&      1.2526&  0.0009&  0.1547&  0.0007&      7.3751 \\
    2456376.85878275&      1.2577&  0.0009&  0.1493&  0.0007&      7.3807 \\
    2456377.75584570&      1.2518&  0.0010&  0.1550&  0.0008&      7.3756 \\
    2456378.74306280&      1.2529&  0.0010&  0.1557&  0.0008&      7.3669 \\
    2456379.77212101&      1.2500&  0.0010&  0.1531&  0.0008&      7.3723 \\
    2456380.77432459&      1.2533&  0.0009&  0.1541&  0.0007&      7.3763 \\
    2456381.77268400&      1.2538&  0.0008&  0.1546&  0.0007&      7.3736 \\
    2456484.55514629&      1.2572&  0.0010&  0.1507&  0.0008&      7.3752 \\
    2456485.56989639&      1.2592&  0.0010&  0.1541&  0.0008&      7.3797 \\
    2456487.56527760&      1.2576&  0.0009&  0.1577&  0.0007&      7.3759 \\
    2456488.54822341&      1.2590&  0.0010&  0.1502&  0.0007&      7.3806 \\
    2456489.53769598&      1.2570&  0.0011&  0.1527&  0.0009&      7.3827 \\
    2456708.80804387&      1.2569&  0.0011&  0.1514&  0.0008&      7.3740 \\
    2456709.82927988&      1.2542&  0.0014&  0.1492&  0.0010&      7.3770 \\
    2456710.82693680&      1.2588&  0.0012&  0.1527&  0.0010&      7.3747 \\
    2456711.80101463&      1.2571&  0.0013&  0.1527&  0.0010&      7.3711 \\
    2456850.53875184&      1.2570&  0.0014&  0.1418&  0.0010&      7.3771 \\
    2456851.59721308&      1.2559&  0.0015&  0.1478&  0.0011&      7.3842 \\
    2456852.57164175&      1.2564&  0.0014&  0.1550&  0.0010&      7.3805 \\
    2456855.58294963&      1.2577&  0.0013&  0.1457&  0.0009&      7.3857 \\
    2456856.55065259&      1.2584&  0.0010&  0.1536&  0.0008&      7.3762 \\
    2457025.79937601&      1.2556&  0.0013&  0.1503&  0.0010&      7.3805 \\
    2457025.85879076&      1.2542&  0.0011&  0.1558&  0.0009&      7.3854 \\
    2457026.86222108&      1.2540&  0.0010&  0.1543&  0.0008&      7.3705 \\
    2457028.85867790&      1.2558&  0.0012&  0.1536&  0.0009&      7.3814 \\
    2457226.56087032&      1.2673&  0.0011&  0.1509&  0.0009&      7.3994 \\
    2457227.53935745&      1.2735&  0.0009&  0.1480&  0.0008&      7.3972 \\
    2457228.50134125&      1.2696&  0.0009&  0.1474&  0.0007&      7.3925 \\
    2457228.58842818&      1.2675&  0.0011&  0.1505&  0.0009&      7.3977 \\
    2457229.48290638&      1.2735&  0.0008&  0.1503&  0.0006&      7.3945 \\
    2457230.49026756&      1.2712&  0.0009&  0.1491&  0.0007&      7.3909 \\
    2457230.61575371&      1.2699&  0.0011&  0.1464&  0.0009&      7.3941 \\
    2457232.51912010&      1.2712&  0.0018&  0.1489&  0.0013&      7.3962 \\
    2457511.82148708&      1.2733&  0.0009&  0.1509&  0.0008&      7.4031 \\
    2457587.57535340&      1.2727&  0.0013&  0.1477&  0.0010&      7.3964 \\
\hline
\end{tabular}
\end{table*}

%

\begin{thebibliography}{}

\bibitem[Anstee \& O'Mara(1995)]{ans95} Anstee, S.~D., \& O'Mara, B.~J.\ 1995, \mnras, 276, 859 

\bibitem[Asplund et 
al.(2005)]{asp05} Asplund, M., Grevesse, N., Sauval, A.~J., Allende Prieto, C., \& Blomme, R.\ 2005, \aap, 431, 693 

\bibitem[Asplund et al.(2009)]{asp09} Asplund, M., Grevesse, N., Sauval, A.~J., \& Scott, P.\ 2009, \araa, 47, 481

\bibitem[Ashwell et al.(2005)]{ash05} Ashwell, J.~F., Jeffries, R.~D., Smalley, B., et al.\ 2005, \mnras, 363, L81 

\bibitem[Barklem 
\& O'Mara(1997)]{bar97} Barklem, P.~S., \& O'Mara, B.~J.\ 1997, \mnras, 290, 102 

\bibitem[Barklem et al.(1998)]{bar98} Barklem, P.~S., O'Mara, 
B.~J., \& Ross, J.~E.\ 1998, \mnras, 296, 1057 

\bibitem[Barklem et al.(2012)]{bar12} Barklem, P.~S., Belyaev, A.~K., Spielfiedel, A., Guitou, M., \& Feautrier, N.\ 2012, \aap, 541, A80 

\bibitem[Baumann et 
al.(2010)]{bau10} Baumann, P., Ram{\'{\i}}rez, I., Mel{\'e}ndez, J., Asplund, M., \& Lind, K.\ 2010, \aap, 519, A87 

\bibitem[Bedell et al.(2014)]{bed14} Bedell, M., 
Mel{\'e}ndez, J., Bean, J.~L., et al.\ 2014, \apj, 795, 23 

\bibitem[Bedell et 
al.(2015)]{bed15} Bedell, M., Mel{\'e}ndez, J., Bean, J.~L., et al.\ 2015, \aap, 581, A34 

\bibitem[Bernstein et al.(2003)]{ber03} Bernstein, R., 
Shectman, S.~A., Gunnels, S.~M., Mochnacki, S., 
\& Athey, A.~E.\ 2003, \procspie, 4841, 1694 

\bibitem[Butler et al.(1993)]{Butler93} Butler, K., Mendoza, C., 
\& Zeippen, C.~J.\ 1993, Journal of Physics B Atomic Molecular Physics, 26, 4409 

\bibitem[Butler 
\& Marcy(1996)]{bm96} Butler, R.~P., \& Marcy, G.~W.\ 1996, \apjl, 464, L153 

\bibitem[Carlos et al.(2016)]{car16}  Carlos, M., Nissen, P.~E., \& Mel{\'e}ndez, J.\ 2016, \aap, 587, A100 

\bibitem[Castelli 
\& Kurucz(2004)]{cas04} Castelli, F., \& Kurucz, R.~L.\ 2004, arXiv:astro-ph/0405087 

\bibitem[Castro et al.(2011)]{cas11} Castro, M., Do Nascimento, J.~D., Jr., Biazzo, K., Mel{\'e}ndez, J., \& de Medeiros, J.~R.\ 2011, \aap, 526, A17 


\bibitem[Chambers(2010)]{cha10} Chambers, J.~E.\ 2010, \apj, 
724, 92 

\bibitem[Charbonnel \& Talon(2005)]{ct05} Charbonnel, C., \& Talon, S.\ 2005, Science, 309, 2189 

\bibitem[Cox(2000)]{cox00} Cox, A.~N.\ 2000, Allen's 
Astrophysical Quantities,  

\bibitem[Cunto 
\& Mendoza(1992)]{Cunto92} Cunto, W., \& Mendoza, C.\ 1992, \rmxaa, 23, 107 

\bibitem[Dawson \& Fabrycky(2010)]{daw10} Dawson, R.~I., \& Fabrycky, D.~C.\ 2010, \apj, 722, 937 

\bibitem[Demarque et al.(2004)]{dem04} Demarque, P., Woo, 
J.-H., Kim, Y.-C., \& Yi, S.~K.\ 2004, \apjs, 155, 667 

\bibitem[Denissenkov(2010)]{den10} Denissenkov, P.~A.\ 2010, \apj, 719, 28 

\bibitem[do Nascimento et al.(2009)]{don09} Do Nascimento, J.~D., Jr., Castro, M., Mel{\'e}ndez, J., Bazot, M., Th{\'e}ado, S., Porto de Mello, G.~F., \& de Medeiros, J.~R.\ 2009, \aap, 501, 687 

\bibitem[dos Santos et al.(2016)]{leo16} dos Santos, L.~A., Mel{\'e}ndez, J., do Nascimento, J.-D., Jr., et al.\ 2016, \aap, 592, A156 

\bibitem[Doyle et al.(2014)]{doy14} Doyle, A.~P., Davies, G.~R., Smalley, B., Chaplin, W.~J., \& Elsworth, Y.\ 2014, \mnras, 444, 3592 

\bibitem[Dumusque et 
al.(2011)]{Dumusque2011} Dumusque, X., Lovis, C., S{\'e}gransan, D., et al.\ 2011, \aap, 535, A55 

\bibitem[Dumusque et al.(2014)]{Dumusque2014} Dumusque, X., Boisse, I., \& Santos, N.~C.\ 2014, \apj, 796, 132 


\bibitem[Fossati et al.(2013)]{Fossati2013} Fossati, L., Ayres, 
T.~R., Haswell, C.~A., et al.\ 2013, \apjl, 766, L20 

\bibitem[Galarza et al.(2016a)]{jhon16a} Galarza, J.~Y., Mel{\'e}ndez, J., Ram{\'{\i}}rez, I., et al.\ 2016a, \aap, 589, A17 

\bibitem[Galarza et al.(2016b)]{jhon16b}  Galarza, J.~Y., Mel{\'e}ndez, J., \& Cohen, J.~G.\ 2016b, \aap, 589, A65 

\bibitem[Gelman \& Rubin(1992)]{gel92} Gelman, A. \& Rubin, D. B.\ 1992, Statistical Science, 7, pp. 457

\bibitem[Gonzalez(1997)]{gon97} Gonzalez, G.\ 1997, \mnras, 285, 403 

\bibitem[Gray(2005)]{gra05} Gray, D.~F.\ 2005, The 
Observation and Analysis of Stellar Photospheres, 3rd Edition. Cambridge, UK:  Cambridge University Press

\bibitem[Haswell et al.(2012)]{Haswell2012} Haswell, C.~A., 
Fossati, L., Ayres, T., et al.\ 2012, \apj, 760, 79 

\bibitem[H{\o}g et 
al.(2000)]{Tycho2000} H{\o}g, E., Fabricius, C., Makarov, V.~V., et al.\ 2000, \aap, 355, L27 

\bibitem[Howard et al.(2011)]{how11} Howard, A.~W., Johnson, J.~A., Marcy, G.~W., et al.\ 2011, \apj, 726, 73 

\bibitem[Kass \& Raftery(1995)]{kass1995} Kass, E. R. \& Raftery, E. A.\ 1995, Journal of the American Statistical Association, 90, 773

\bibitem[Lambert(1993)]{lam93} Lambert, D.~L.\ 1993, Physica 
Scripta Volume T, 47, 186 

\bibitem[Lind et 
al.(2009)]{lin09} Lind, K., Asplund, M., \& Barklem, P.~S.\ 2009, \aap, 503, 541 

\bibitem[Lind et 
al.(2011)]{lin11} Lind, K., Asplund, M., Barklem, P.~S., \& Belyaev, A.~K.\ 2011, \aap, 528, A103 

\bibitem[Lo Curto et al.(2015)]{loc15} Lo Curto, G., Pepe, 
F., Avila, G., et al.\ 2015, The Messenger, 162, 9 

\bibitem[Lovis et al.(2011)]{Lovis2011} Lovis, C., Dumusque, X., 
Santos, N.~C., et al.\ 2011, arXiv:1107.5325 

\bibitem[Mamajek 
\& Hillenbrand(2008)]{Mamajek2008} Mamajek, E.~E., \& Hillenbrand, L.~A.\ 2008, \apj, 687, 1264 

\bibitem[Mayor 
\& Queloz(1995)]{mq95} Mayor, M., \& Queloz, D.\ 1995, \nat, 378, 355

\bibitem[Mayor et al.(2003)]{may03} Mayor, M., Pepe, F., 
Queloz, D., et al.\ 2003, The Messenger, 114, 20 

\bibitem[Marcy 
\& Butler(1996)]{mb96} Marcy, G.~W., \& Butler, R.~P.\ 1996, \apjl, 464, L147

\bibitem[Mel{\'e}ndez et 
al.(2007)]{mel07} Mel{\'e}ndez, M., Bautista, M.~A., \& Badnell, N.~R.\ 2007, \aap, 469, 1203 

\bibitem[Mel{\'e}ndez et al.(2009)]{mel09} Mel{\'e}ndez, J., 
Asplund, M., Gustafsson, B., \& Yong, D.\ 2009, \apjl, 704, L66 

\bibitem[Mel{\'e}ndez et al.(2012)]{mel12} Mel{\'e}ndez, J., Bergemann, M., Cohen, J.~G., et al.\ 2012, \aap, 543, A29 

\bibitem[Mel{\'e}ndez et al.(2014a)]{mel14a} Mel{\'e}ndez, J., Ram{\'{\i}}rez, I., Karakas, A.~I., et al.\ 2014a, \apj, 791, 14 

\bibitem[Mel{\'e}ndez et al.(2014b)]{mel14b} Mel{\'e}ndez, J., Schirbel, L., Monroe, T.~R., et al.\ 2014b, \aap, 567, L3 

\bibitem[Monroe et al.(2013)]{mon13} Monroe, T.~R., Mel{\'e}ndez, J., Ram{\'{\i}}rez, I., et al.\ 2013, \apjl, 774, L32 

\bibitem[Montalb{\'a}n \& Rebolo(2002)]{mon02} Montalb{\'a}n, J., \& Rebolo, R.\ 2002, \aap, 386, 1039 

\bibitem[Murray \& Chaboyer(2002)]{mur02} Murray, N., \& Chaboyer, B.\ 2002, \apj, 566, 442 

\bibitem[Nissen(2015)]{nis15} Nissen, P.~E.\ 2015, \aap, 579, A52 

\bibitem[{\"O}nehag et al.(2011)]{one11} {\"O}nehag, A., Korn, A., Gustafsson, B., Stempels, E., \& Vandenberg, D.~A.\ 2011, \aap, 528, A85 

\bibitem[Osorio et 
al.(2015)]{oso15} Osorio, Y., Barklem, P.~S., Lind, K., et al.\ 2015, \aap, 579, A53 

\bibitem[Pepe et al.(2011)]{Pepe2011} Pepe, F., Lovis, C., S{\'e}gransan, D., et al.\ 2011, \aap, 534, A58 

\bibitem[Ram{\'{\i}}rez et 
al.(2007)]{ram07} Ram{\'{\i}}rez, I., Allende Prieto, C., \& Lambert, D.~L.\ 2007, \aap, 465, 271 

\bibitem[Ram{\'{\i}}rez et 
al.(2009)]{ram09} Ram{\'{\i}}rez, I., Mel{\'e}ndez, J., \& Asplund, M.\ 2009, \aap, 508, L17 

\bibitem[Ram{\'{\i}}rez et al.(2011)]{ram11} Ram{\'{\i}}rez, 
I., Mel{\'e}ndez, J., Cornejo, D., Roederer, I.~U., \& Fish, J.~R.\ 2011, \apj, 740, 76 

\bibitem[Ram{\'{\i}}rez et al.(2012)]{ram12} Ram{\'{\i}}rez, I., Mel{\'e}ndez, J., \& Chanam{\'e}, J.\ 2012, \apj, 757, 164 

\bibitem[Ram{\'{\i}}rez et al.(2013)]{ram13} Ram{\'{\i}}rez, 
I., Allende Prieto, C., \& Lambert, D.~L.\ 2013, \apj, 764, 78 

\bibitem[Ram{\'{\i}}rez et 
al.(2014)]{ram14} Ram{\'{\i}}rez, I., Mel{\'e}ndez, J., Bean, J., et al.\ 2014, \aap, 572, AA48 

\bibitem[Ram{\'{\i}}rez et al.(2015)]{ram15} Ram{\'{\i}}rez, I., Khanal, S., Aleo, P., et al.\ 2015, \apj, 808, 13 

\bibitem[Samson 
\& Berrington(2001)]{sam01} Samson, A.~M., \& Berrington, K.~A.\ 2001, Atomic Data and Nuclear Data Tables, 77, 87 

\bibitem[Sandquist et al.(2002)]{san02} Sandquist, E.~L., 
Dokter, J.~J., Lin, D.~N.~C., \& Mardling, R.~A.\ 2002, \apj, 572, 1012 

\bibitem[Santos et 
al.(2004)]{san04} Santos, N.~C., Bouchy, F., Mayor, M., et al.\ 2004, \aap, 426, L19 

\bibitem[Shi et al.(2014)]{shi14} Shi, J.~R., Gehren, T., 
Zeng, J.~L., Mashonkina, L., \& Zhao, G.\ 2014, \apj, 782, 80 

\bibitem[Siess et 
al.(2000)]{sie00} Siess, L., Dufour, E., \& Forestini, M.\ 2000, \aap, 358, 593 

\bibitem[Sneden(1973)]{sne73} Sneden, C.~A.\ 1973, Ph.D.~Thesis,  

\bibitem[Spina et al.(2015)]{spi15} Spina, L., Palla, F., Randich, S., et al.\ 2015, \aap, 582, L6 

\bibitem[Spina et al.(2016)]{spi16} Spina, L., Mel{\'e}ndez, J., \& Ram{\'{\i}}rez, I.\ 2016, \aap, 585, A152 

\bibitem[Teske et al.(2016)]{tes16} Teske, J.~K., Khanal, S., \& Ram{\'{\i}}rez, I.\ 2016, \apj, 819, 19 

\bibitem[Th{\'e}ado \& Vauclair(2012)]{the12} Th{\'e}ado, S., \& Vauclair, S.\ 2012, \apj, 744, 123 

\bibitem[Tognelli et al.(2016)]{tog16} Tognelli, E., Prada Moroni, P.~G., \& Degl'Innocenti, S.\ 2016, \mnras, 460, 3888 

\bibitem[Tucci Maia et al.(2014)]{tuc14} Tucci Maia, M., 
Mel{\'e}ndez, J., \& Ram{\'{\i}}rez, I.\ 2014, \apjl, 790, LL25

\bibitem[Tucci Maia et al.(2016)]{tuc16} Tucci Maia, M., Ram{\'{\i}}rez, I., Mel{\'e}ndez, J., et al.\ 2016, \aap, 590, A32 

\bibitem[Xiong \& Deng(2009)]{xd09} Xiong, D.~R., \& Deng, L.\ 2009, \mnras, 395, 2013 

\bibitem[Yan et al.(2015)]{yan15} Yan, H.~L., Shi, J.~R., 
\& Zhao, G.\ 2015, \apj, 802, 36 

\bibitem[Zakamska et al.(2011)]{Zakamska2011} Zakamska, N.~L., Pan, 
M., \& Ford, E.~B.\ 2011, \mnras, 410, 1895 

\bibitem[Zatsarinny et al.(2009)]{zat09} Zatsarinny, O., 
Bartschat, K., Gedeon, S., et al.\ 2009, \pra, 79, 052709 

\bibitem[Zechmeister \& K{\"u}rster(2009)]{Zechmeister2009} Zechmeister, M., K{\"u}rster, M.\ 2009, \aap, 496, 577 

\end{thebibliography}
%

\Online

\begin{appendix} 
\section{NLTE corrections for Mg and Ca}

Our Mg and Ca atom models for NLTE calculations are unpublished,
 although our Mg model is somewhat similar to the Mg atom model recently published by \cite{oso15}. A brief description
of our Mg and Ca models is given below.
Radiative data were downloaded from the Kurucz (http://www.pmp.uni-hannover.de/cgi-bin/ssi/test/kurucz/sekur.html) 
and TOPbase \citep{Cunto92} web interfaces in July 2011. When necessary, the atoms were supplemented with hydrogenic data. 
The TOPbase reference for Mg I is \cite{Butler93}, for Mg II it is ``K.T. Taylor, to be published'', for CaI and CaII it is ``H.E. Saraph \& P.J. Storey, to be published''.
We included 220 energy levels for Mg, complete up to $n$ = 80 for Mg I and up to $n$ = 10 for Mg II. 
Fine structure is resolved up to $l \leq$ 3 ($s,p,d,f$) and $n \leq$19 in Mg I; $n$ = 20-80 are collapsed super levels.  
For Ca we included 256 energy levels, with identical coverage as for Mg.
For Mg and Ca, 1390 and 1961 b-b radiative transitions were included, respectively; 
$f$-values are taken from TOPbase preferentially, otherwise from the Kurucz database. 
Collisional broadening due to collisions with neutral H is based on the
ABO theory \citep{ans95,bar97,bar98}  when possible ($\sim$15 lines for each atom). Stark broadening is neglected.
Photoionisation cross-sections are from TOPbase for both ionisation stages. 
Electron excitation and ionisation were taken from \cite{cox00}. Cross-sections for forbidden transitions were computed 
assuming f=0.5, and when available replaced with \cite{zat09} for Mg and \cite{sam01} and \cite{mel07} for Ca. 
For hydrogen collisions we adopted Sh=0.1 \citep{lam93}, and replaced with \cite{bar12} when available for Mg.

\end{appendix}

\end{document}